\documentclass[%
 aip, pop,
 amsmath,amssymb,
 reprint,%
]{revtex4-1}

\usepackage{graphicx}
\usepackage{dcolumn}
\usepackage{bm}
\usepackage[mathlines]{lineno}

\usepackage[utf8]{inputenc}
\usepackage[T1]{fontenc}
\usepackage{mathptmx}
\usepackage{etoolbox}
\usepackage{orcidlink}   

\makeatletter
\def\@email#1#2{%
 \endgroup
 \patchcmd{\titleblock@produce}
  {\frontmatter@RRAPformat}
  {\frontmatter@RRAPformat{\produce@RRAP{*#1\href{mailto:#2}{#2}}}\frontmatter@RRAPformat}
  {}{}
}%
\makeatother
\begin{document}

\preprint{AIP/123-QED}

\title[Investigation of resonant layer response in electron viscosity regime]{Investigation of resonant layer response in electron viscosity regime}
\author{Yeongsun Lee\orcidlink{0000-0003-4474-416X}}
\thanks{These authors contributed equally to this work.}
\affiliation{Seoul National University, Seoul, South Korea}
\affiliation{Nuclear Research Institute for Future Technology and Policy, Seoul National University, Seoul, South Korea}

\author{Jace Waybright\orcidlink{0000-0001-9573-061X}}
\thanks{These authors contributed equally to this work.}
\affiliation{Princeton Plasma Physics Laboratory}

\author{Jong-Kyu Park\orcidlink{0000-0003-2419-8667}$^*$}%
\email{jkpark@snu.ac.kr}
\affiliation{Seoul National University, Seoul, South Korea}
\affiliation{Princeton Plasma Physics Laboratory}

\date{\today}

\begin{abstract}
We present a supplementary study of previous work in Waybright and Park [Phys. Plasmas \textbf{31}, 022502 (2024)] which demonstrates a substantial effect of electron viscosity on the resonant layer response to non-axisymmetric magnetic perturbations. A main refinement is to include a curl element of electron viscosity in the generalized Ohm's law. The refinement reveals a resonant layer response in the Electron Viscosity (EV) regime corresponding to slowly rotating and highly viscous plasmas.
\end{abstract}

\maketitle

\section{Introduction}
In a tokamak, there are non-axisymmetric magnetic perturbations such as intrinsic error fields (EF) and externally applied resonant magnetic perturbations (RMP). A small amplitude of resonant perturbations can be significantly amplified by resonance on a rational surface, driving magnetic reconnection and exerting braking force, if the resonance condition $\vec{k} \cdot \vec{B}=0$ is satisfied \cite{Furth1963PoF, Hahm1985PoF, Fitzpatrick1993NF}. In-depth understanding of resonant layer responses is of great importance in ensuring stable and high-performance operations in tokamaks such as the EF correction and RMP-induced Edge Localized Mode controls in H-mode plasmas \cite{Evans2006NP, Park2018NP, Kim2024NC, Yang2024Nat}. 

Investigation of resonant layer response has inherent complexity \cite{Waelbroeck2003PoP, Waelbroeck2012NF,Cole2006PoP, Cole2008PoP, Fitzpatrick2022PoP, Fitzpatrick2023PoP}. For instance, at least ten analytic simplifications were found to be possible under the two-fluid drift-MHD formalism \cite{Cole2006PoP}. Recently, the analytic technique of the two-layer match \cite{Cole2006PoP} was further advanced by accounting for electron viscosity in the parallel generalized Ohm's law \cite{Waybright2024PoP}, where the primary contribution to the parallel Ohm's law shifts from resistivity to electron viscosity. This previous study provides the physical motivation of this work, which is the same transition of dominant processes in the curl of the generalized Ohm's law may take place when the electron physics plays the key role in determining $\Delta$; otherwise, the only important effect is the transition in the parallel Ohm's law. Therefore, we present a supplementary study by incorporating the curl of the generalized Ohm's law, clarifying the resonant layer response in the electron viscosity regime, and verifying an analytic solution using the Slab LAYER code (SLAYER) \cite{Park2022PoP, Waybright2024PoP, Lee2024NF}.

\section{Layer physics}
The two-fluid drift-MHD layer equations are reduced into the so-called four-field equations\cite{Fitzpatrick2005PoP, Hazeltine1985PoF},
\begin{equation}
    i(Q - Q_{e}) \tilde\psi = iX(\tilde\phi - \tilde Z) + \frac{d^2 \tilde\psi}{dX^2} - (1+\tau) P_e (\frac{d^2 \tilde V_z}{dX^2} + \frac{D^2}{c_\beta^2}\frac{d^4 \tilde\psi}{dX^4}), \label{eq:paraohm_c}
\end{equation}
\begin{equation}
\begin{split}
    iQ \tilde Z = &iQ_{\ast e} \tilde\phi + iD^2 X \frac{d^2 \tilde\psi}{dX^2} +ic_\beta^2 X \tilde V_z \\
    &+ (c_\beta^2 + (1-c_\beta^2)K) \frac{d^2 \tilde Z}{dX^2} + P_e D^2 (\frac{d^4 \tilde \phi}{dX^4}-\frac{d^4 \tilde Z}{dX^4}), \label{eq:curlohm_c}
\end{split}
\end{equation}
\begin{equation}
    i(Q-Q_{\ast i}) \frac{d^2 \tilde \phi}{dX^2} = iX\frac{d^2 \tilde\psi}{dX^2} + P(\frac{d^4 \tilde \phi}{dX^4}+\tau\frac{d^4 \tilde Z}{dX^4}) + P_e (\frac{d^4 \tilde \phi}{dX^4}-\frac{d^4 \tilde Z}{dX^4}), \label{eq:curlion_c}
\end{equation}
\begin{equation}
    iQ \tilde V_z = -iQ_{\ast e} \tilde\psi + iX\tilde Z + P \frac{d^2 \tilde V_z}{dX^2} +  P_e \frac{d^2 \tilde V_z}{dX^2} + P_e \frac{D^2}{c_\beta^2} \frac{d^4 \tilde\psi}{dX^4},\label{eq:paraion_c}
\end{equation}
where the nomenclatures are borrowed from 
Ref. \cite{Waybright2024PoP}. Equations \ref{eq:paraohm_c} and \ref{eq:curlohm_c} are deduced from the generalized ohm's law by taking $\hat{z}\cdot$ and $\hat{z}\cdot \vec{\nabla}\times$, respectively; similarly, equations \ref{eq:curlion_c} and \ref{eq:paraion_c} are from the ion momentum equation by taking $\hat{z}\cdot \vec{\nabla}\times$ and $\hat{z}\cdot$, respectively. We neglected the ion screening term $c_\beta^2 X \tilde V_z$ \cite{Lee2024NF} by limiting our interest to the low toroidal $\beta$ limit. The parallel flow term in electron viscosity part $(1+\tau) P_e \frac{d^2 \tilde V_z}{dX^2}$ is also ignored since its role is expected to be less significant than $(1+\tau) P_e \frac{D^2}{c_\beta^2}\frac{d^4 \tilde\psi}{dX^4}$ in parametric domains with strong electron viscosity. Then, the layer response can be accounted for by the three-field equations (Eqns. \ref{eq:paraohm_c}-\ref{eq:curlion_c}).

The three-field equations in phase space after taking a Fourier transformation become, 
\begin{align}
    i(Q-Q_e) \bar{\psi} &= \frac{d (\bar{\phi} - \bar{Z})}{dp}-p^2 \bar{\psi} - p^4 P_e \frac{D^2}{c_\beta^2}(1+\tau) \bar{\psi}, \label{eq:para_ohms_law}\\
    iQ \bar{Z} - iQ_e \bar{\phi} &= -D^2 \frac{d^2 (p^2 \bar{\psi})}{dp^2} - c_\beta^2 p^2 \bar{Z} + P_e D^2 p^4 (\bar{\phi} - \bar{Z}), \label{eq:curl_ohms_law}\\
    i(Q-Q_i) p^2 \bar{\phi} &= \frac{d(p^2 \bar{\psi})}{dp} - Pp^4 (\bar{\phi} + \tau \bar{Z}) - P_e p^4 (\bar{\phi} - \bar{Z}) \label{eq:curl_ion_law}.
\end{align}
These equations can be reduced into the 2nd order ODE with the variable $\bar{Y} \equiv \bar{\phi} - \bar{Z}$,
\begin{equation}
    \frac{d}{dp} \Big[ \frac{p^2}{\underbrace{\frac{D^2(\tau+1)P_e}{c_\beta^2}p^4}_{\text{parallel electron viscosity}} + p^2+i(Q-Q_e)} \frac{d\bar{Y}}{dp}\Big] - p^2G(p) \bar{Y} = 0, \label{eq:reduced_layer}
\end{equation}
where $G(p)$ has a correction from inclusion of electron viscosity curl in Eqn. \ref{eq:curl_ohms_law} with approximation $P+P_e \approx P$
\begin{widetext}
\begin{equation}
    G(p) = \frac{\overbrace{PP_eD^2(\tau+1)p^6+ \Big(iP_eD^2(Q-Q_i)}^{\text{correction from curls of electron viscosity}} + P c_\beta^2 \Big) p^4 + i (c_\beta^2 + P) (Q-Q_i) p^2 - Q(Q-Q_i)}{PD^2(\tau+1) p^4 + \Big( i(Q-Q_i)D^2 + c_\beta^2 \Big)p^2 + i (Q-Q_e)}.
\end{equation}    
\end{widetext}
In nonconstant-$\psi$ regimes, effects of electron viscosity are negligible \cite{Waybright2024PoP}. Hence, we only focus on the constant-$\psi$ regimes where $\max(\frac{D^2(\tau+1)P_e}{c_\beta^2}p^4, p^2) \gg Q$ is met. When the parallel electron viscosity is weak, i.e. $\frac{D^2(\tau+1)P_e}{c_\beta^2}p^4\ll p^2$, the $\hat{\Delta}$ is obtained using the standard two-layer matching technique \cite{Cole2006PoP}, whereas for $p^2 \ll \frac{D^2(\tau+1)P_e}{c_\beta^2}p^4$, the higher-order matching \cite{Waybright2024PoP} is employed.

Suppose that $Q\gg c_\beta^2 p^2$ and $Q \ll P p^2$. Equation \ref{eq:reduced_layer} reduces to
\begin{equation}
    \frac{d}{dp} \Big[ \frac{1}{\frac{D^2(\tau+1) P_e}{c_\beta^2}p^2 + 1} \frac{d \bar{Y}}{dp} \Big] - Pp^4 \frac{P_eD^2(\tau+1)p^4 + i(Q-Q_i)}{PD^2(\tau+1)p^4 + i(Q-Q_e)} \bar{Y} \approx 0.
\end{equation}
This reduction is equivalent to writing Equations \ref{eq:para_ohms_law}-\ref{eq:curl_ion_law} as
\begin{align}
    0 &= \frac{d (\bar{\phi} - \bar{Z})}{dp}-p^2 \bar{\psi} - p^4 P_e \frac{D^2}{c_\beta^2}(1+\tau) \bar{\psi}, \\
    iQ \bar{Z} - iQ_e \bar{\phi} &= -D^2 \frac{d^2 (p^2 \bar{\psi})}{dp^2}\\
    i(Q-Q_i) p^2 \bar{\phi} &= \frac{d(p^2 \bar{\psi})}{dp} - Pp^4 (\bar{\phi} + \tau \bar{Z}) - P_e p^4 (\bar{\phi} - \bar{Z}).
\end{align}
That is, a primary correction originates from accounting for electron viscosity in curl of ion momentum equation compared to Ref. \cite{Waybright2024PoP}. Therefore, one can expect that the correction is trivial due to a negligible electron mass, i.e. $P_e/P \ll 1$. In the limit $Q \gg PD^2p^4$, called the first visco-resistive (VRi) regime, Equation \ref{eq:reduced_layer} reduces to
\begin{equation}
    \frac{d^2\bar{Y}}{dp^2} - \frac{Q-Q_i}{Q-Q_e} Pp^4 \bar{Y} \approx 0,
\end{equation}
and results in
\begin{equation}
    \hat{\Delta}_{VRi} = \frac{6^{2/3} i \pi \Gamma(5/6)(Q-Q_i)^{1/6}(Q-Q_e)^{5/6} P^{1/6}}{\Gamma(1/6)} .
\end{equation}
In the limit $Q \ll PD^2p^4$ and $\frac{D^2(\tau+1)P_e}{c_\beta^2} p^2 \ll 1$, called the first semicollisional (SCi) regime, Equation \ref{eq:reduced_layer} reduces to
\begin{equation}
    \frac{d^2\bar{Y}}{dp^2} - i \frac{Q-Q_i}{D^2(\tau+1)} \bar{Y} \approx 0,
\end{equation}
and results in
\begin{equation}
    \hat{\Delta}_{SCi} =  \frac{i^{3/2}\pi(Q-Q_i)^{1/2}(Q-Q_e)}{D(\tau+1)^{1/2}}.
\end{equation}
In the limit $\frac{D^2(\tau+1)P_e}{c_\beta^2} p^2 \gg 1$ and $Q \gg P_eD^2p^4$, the first semicollisional regime with a correction from electron viscosity (SCiPe), Equation \ref{eq:reduced_layer} reduces to
\begin{equation}
    \frac{d}{dp} \Big[ \frac{1}{p^2} \frac{d\bar{Y}}{dp}\Big] - i P_e\frac{Q-Q_i}{c_\beta^2} \bar{Y} \approx 0,
\end{equation}
and results in
\begin{equation}
    \hat{\Delta}_{SCiPe} = \frac{3\Gamma(1/4) i^{7/4} \pi c_\beta^{1/2}(Q-Q_e)(Q-Q_i)^{3/4} }{8  \Gamma(7/4)D^2(\tau+1)P_e^{1/4}}.
\end{equation}
Note that a key role played by electron viscosity comes from parallel element of generalized Ohm's law, which is consistent with our expectation.

Suppose that $Q \ll c_\beta^2 p^2$ and $Q \ll Pp^2$. Equation \ref{eq:reduced_layer} reduces to
\begin{equation}
    \frac{d}{dp} \Big[ \frac{1}{\frac{D^2(\tau+1) P_e}{c_\beta^2}p^2 + 1} \frac{d \bar{Y}}{dp} \Big] - Pp^4 \frac{P_eD^2(\tau+1)p^2 + c_\beta^2}{PD^2(\tau+1)p^2 + c_\beta^2} \bar{Y} \approx 0.
\end{equation}
This reduction is equivalent to writing Equations \ref{eq:para_ohms_law}-\ref{eq:curl_ion_law} as
\begin{align}
    0 &= \frac{d (\bar{\phi} - \bar{Z})}{dp}-p^2 \bar{\psi} - p^4 P_e \frac{D^2}{c_\beta^2}(1+\tau) \bar{\psi}, \\
    0 &= -D^2 \frac{d^2 (p^2 \bar{\psi})}{dp^2} - c_\beta^2 p^2 \bar{Z} + P_e D^2 p^4 (\bar{\phi} - \bar{Z}), \\
    0 &= \frac{d(p^2 \bar{\psi})}{dp} - Pp^4 (\bar{\phi} + \tau \bar{Z}).
\end{align}
That is, a primary correction originates from accounting for electron viscosity in the curl of generalized Ohm's law compared to Ref. \cite{Waybright2024PoP}. Therefore, one can expect that the correction is nontrivial in this case. In the limit $c_\beta^2 \gg PD^2p^2$, called the second visco-resistive (VRii) regime, Equation \ref{eq:reduced_layer} reduces to
\begin{equation}
    \frac{d^2\bar{Y}}{dp^2} - Pp^4 \bar{Y} \approx 0,
\end{equation}
and results in
\begin{equation}
    \hat{\Delta}_{VRii} = \frac{6^{2/3} i\pi \Gamma(5/6) (Q-Q_e) P^{1/6}}{\Gamma(1/6)}.
\end{equation}
In the limit $c_\beta^2 \ll PD^2p^2$ and $\frac{D^2(\tau+1)P_e}{c_\beta^2} p^2 \ll 1$, called the first Hall-resistive (HRi) regime, Equation \ref{eq:reduced_layer} reduces to
\begin{equation}
    \frac{d^2\bar{Y}}{dp^2} - \frac{c_\beta^2}{D^2(\tau+1)}p^2 \bar{Y} \approx 0,
\end{equation}
and results in
\begin{equation}
    \hat{\Delta}_{HRi} = \frac{2i \pi \Gamma(3/4)c_\beta^{1/2}(Q-Q_e)}{\Gamma(1/4)D^{1/2}(\tau+1)^{1/4}} .
\end{equation}
In the limit $\frac{D^2(\tau+1)P_e}{c_\beta^2} p^2 \gg 1$, which we shall refer to as the electron-viscosity (EV) regime, it reduces to
\begin{equation}
    \frac{d}{dp} \Big[ \frac{1}{p^2} \frac{d\bar{Y}}{dp}\Big] - \frac{D^2(\tau+1)P_e^2}{c_\beta^2} p^4 \bar{Y} \approx 0
\end{equation}
and results in
\begin{equation}
    \hat{\Delta}_{EV} = \frac{3 i\pi\Gamma(5/8)c_\beta^{5/4}(Q-Q_e)}{8^{3/4}\Gamma(11/8)D^{5/4}(\tau+1)^{5/8}P_e^{1/4}}.
\end{equation}
Note that the curl element of electron viscosity in the generalized Ohm's law has a comparable impact on the layer response compared to the parallel element, which indeed aligns with our expectation.

\begin{figure}
    \centering
    \includegraphics[width=\linewidth]{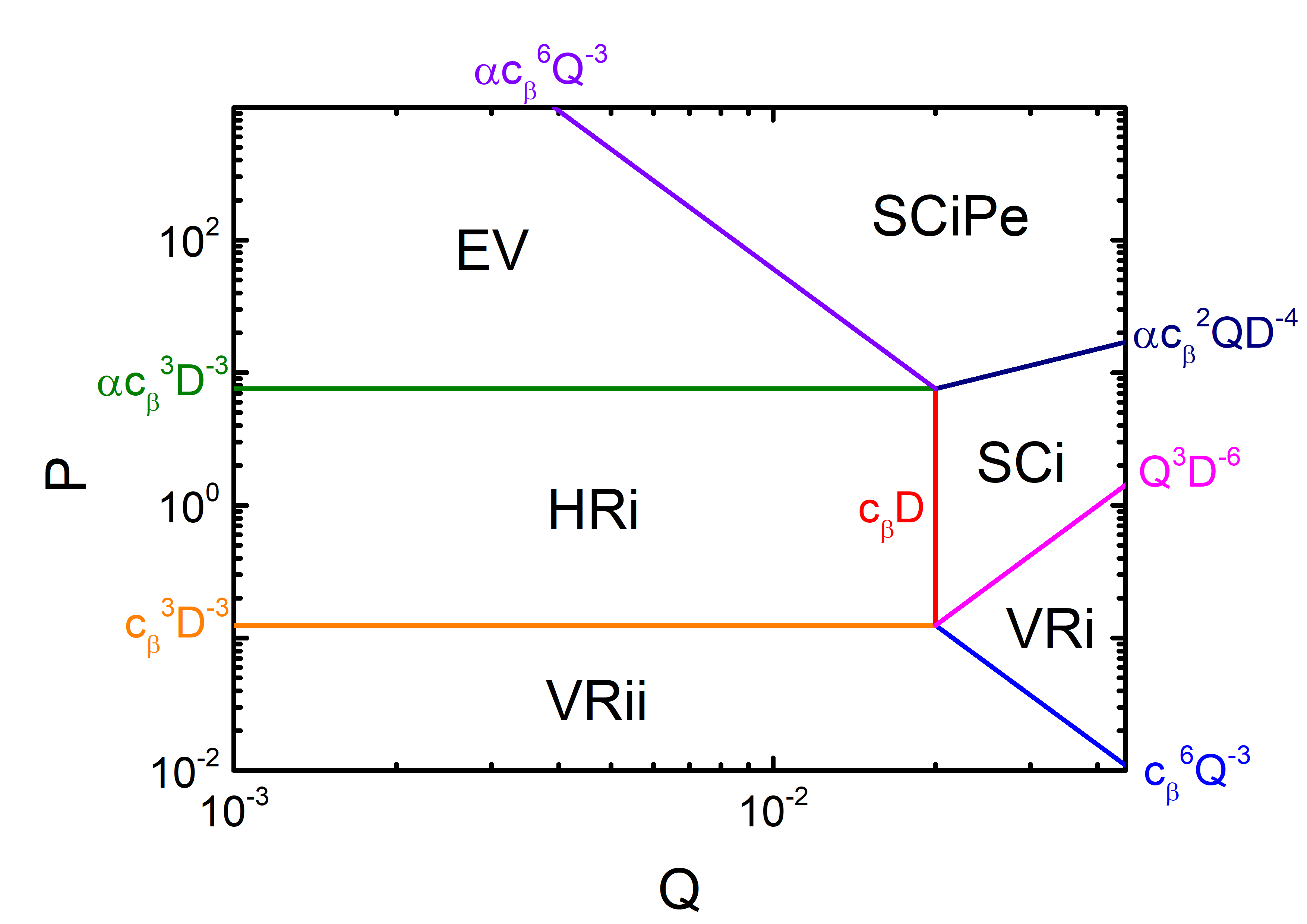}
    \caption{An analytic regime map in $Q-P$ space illustrating the classification of regimes. $c_\beta=0.1$, $D=0.2$, $Q_e=-Q_i=Q/2$ and $\alpha = P/P_e$.}
    \label{fig:analytic_regime_map}
\end{figure}

The distinct regimes have been identified relying not only on the parameters but also on the $p$-variable. By substituting the layer width $p_*$\cite{Cole2006PoP, Waybright2024PoP} into $p$, the regimes are mapped onto $Q-P$ space in Figure \ref{fig:analytic_regime_map}. Regime boundaries do not imply sharp transitions; instead, they serve as indicative lines far from which the asymptotic approximations are expected to hold.

We verify the analytic asymptotes using SLAYER \cite{Park2022PoP} which solves layer equations (Eqns. \ref{eq:paraohm_c} - \ref{eq:paraion_c}), neglecting the ion screening term, by applying the higher order Riccati transformation method adapted to the layer problem \cite{Lee2024NF}.

\begin{figure}
    \centering
    \includegraphics[width=\linewidth]{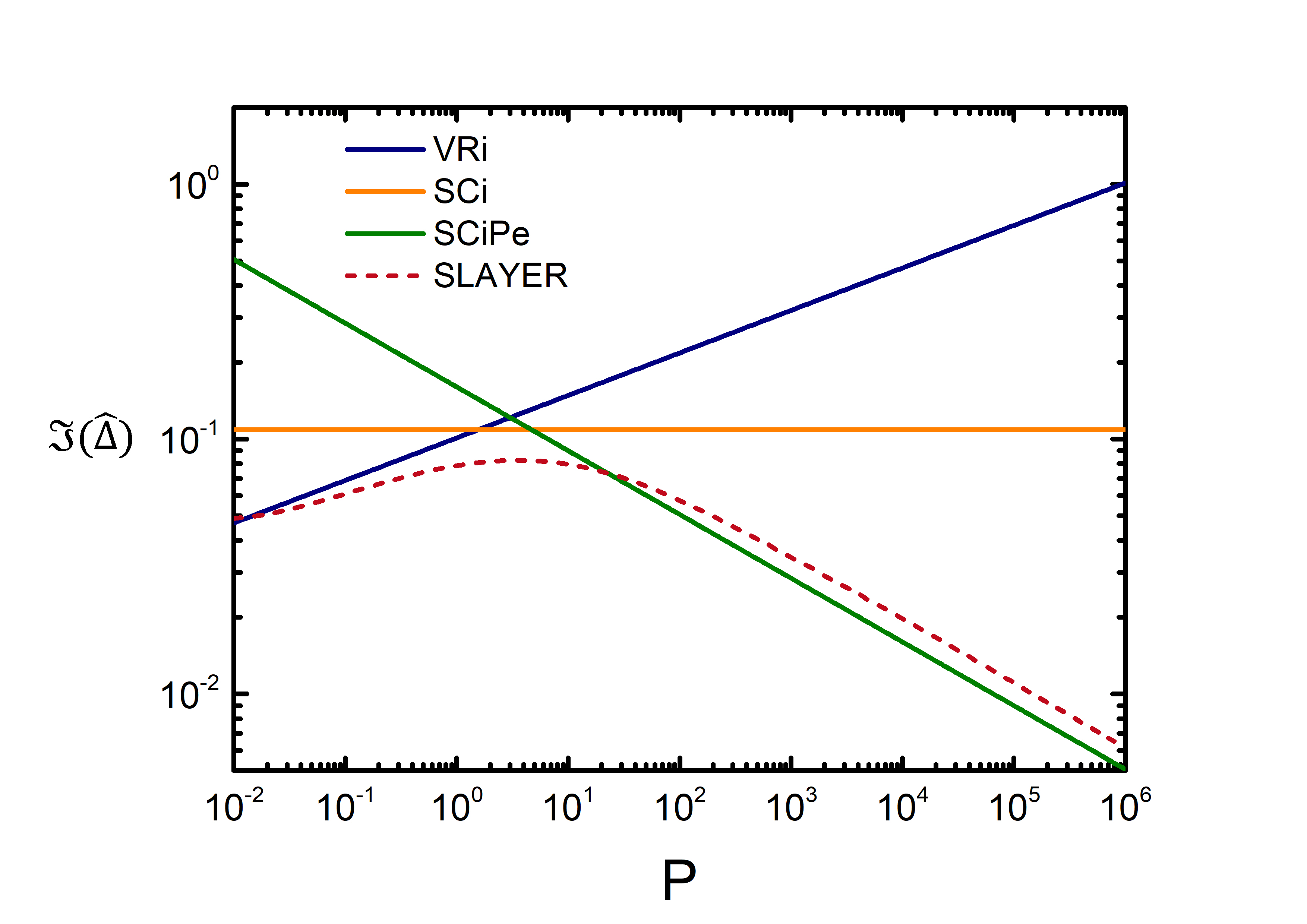}
    \caption{Transition from the VRi to SCi to SCiPe regime. $Q=0.08$, $Q_e=Q/2$, $Q_i=-Q/2$, $D=0.2$, $c_\beta=0.1$ and $P_e/P = 0.0165$.}
    \label{fig:transition_VRi_SCi_SCiPe}
\end{figure}

Transition from the VRi to SCi to SCiPe regime is demonstrated in the $P$-space in Fig. \ref{fig:transition_VRi_SCi_SCiPe}. As $P$ increases, the numerical solutions show a great agreement with the analytic solutions, from the VRi to SCi to SCiPe regime in order. 

\begin{figure}
    \centering
    \includegraphics[width=\linewidth]{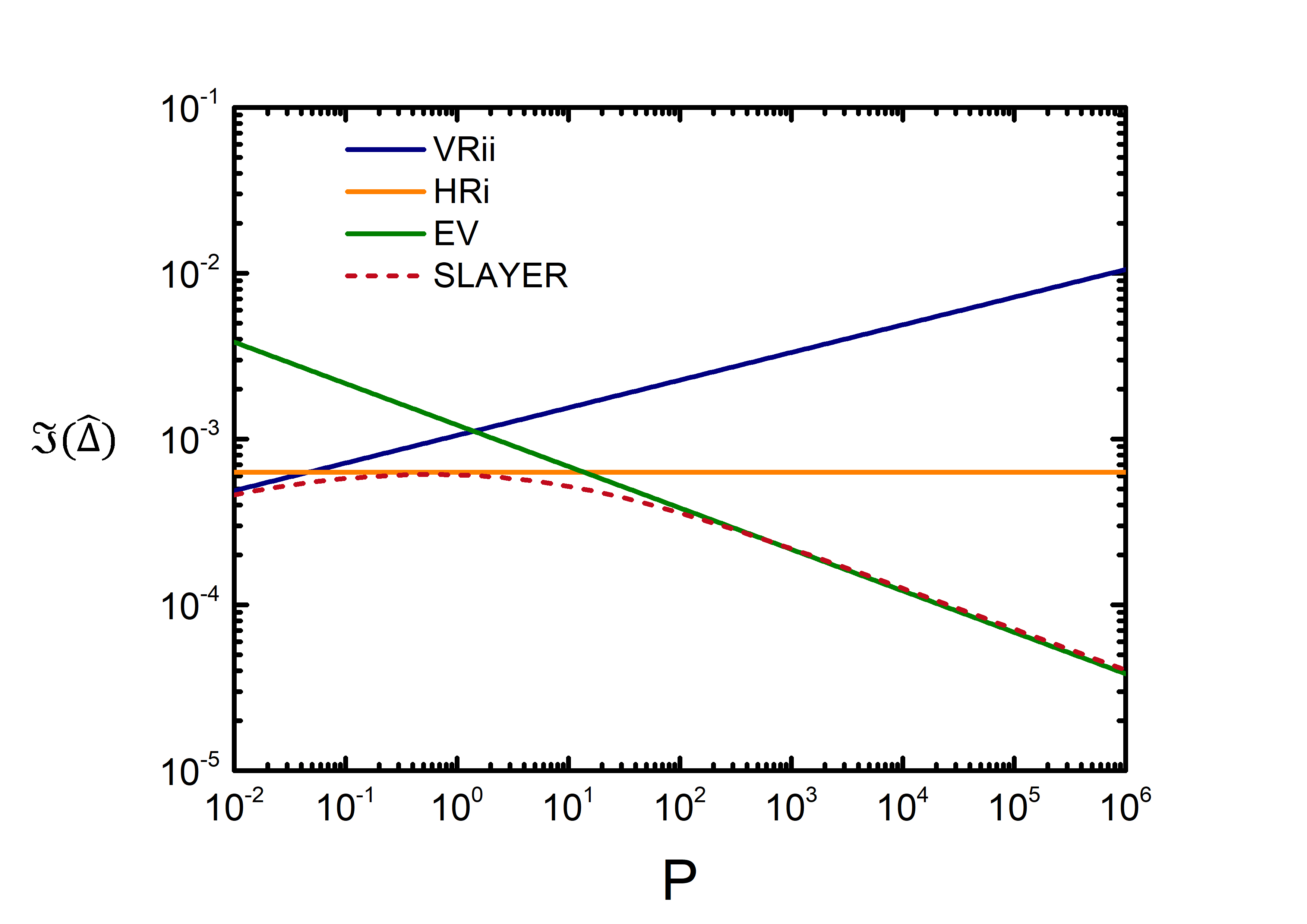}
    \caption{Transition from the VRii to HRi to EV regime. $Q=10^{-3}$, $Q_e=Q/2$, $Q_i=-Q/2$, $D=0.2$, $c_\beta=0.1$ and $P_e/P = 0.0165$.}
    \label{fig:transition_VRii_HRi_EV}
\end{figure}

Transition from the VRii to HRi to EV regime is demonstrated in the $P$-space in Fig. \ref{fig:transition_VRii_HRi_EV}. As $P$ increases, the numerical solutions show a remarkable agreement with the analytic solutions, from the VRii to HRi to EV regime in order. As discussed, access to the EV regime is allowed when electron viscosity is accounted for in the curl of Ohm's law since the condition $P_e D^2(\tau+1)p^4 \gg c_\beta^2 p^2$ required for dominant electron viscosity is simultaneously satisfied in both the parallel and curl part of Ohm's law.

\section{Field penetration scaling in EV regime}
Suppose the field penetration threshold in a large aspect ratio, low beta, ohmically heated tokamak plasma \cite{Cole2006PoP}. In the EV regime, applying the standard scaling technique \cite{Cole2006PoP, Fitzpatrick2022PoP, Waybright2024PoP} yields the field penetration scaling law
\begin{equation}
    \Big[ \frac{b_r(r_s)}{B_\phi} \Big]_{critEV} \sim \rho_\ast^{3/8} \nu_\ast^{1/2} \beta^{3/16},
\end{equation}
which in engineering parameters is
\begin{equation}
    \Big[ \frac{b_r(r_s)}{B_\phi} \Big]_{critEV} \sim n_e^{11/16} T_e^{-5/8} R_0^{1/8} B_\phi^{-3/4}.
\end{equation}
Although the density scaling coefficient becomes weaker than the previous expectation\cite{Waybright2024PoP}-$\Big[ \frac{b_r(r_s)}{B_\phi} \Big]_{critHRiHe} \sim n_e R_0^{-1/2} B_\phi^{-2}$-, it is still stronger than that in the limit of no electron viscosity\cite{Cole2006PoP}-$\Big[ \frac{b_r(r_s)}{B_\phi} \Big]_{critHRi} \sim n_e^{1/4} T_e^{1/8} R_0^{-1} B_\phi^{-5/4}$.

It is worth noting that the temperature scaling coefficient carries a negative sign. When considered alongside the accessibility of the EV regime, characterized by low rotation and low electron temperature ($P \sim 1/\beta $ according to classical estimation \cite{Cole2006PoP}), this negative temperature scaling implies stronger shielding via electron viscosity nearby the pedestal foot under the RMP effect than predicted by the HRi regime. Hence, this scaling can provide guidance on the RMP optimization, especially aiming at edge-localized RMP \cite{Yang2024Nat}.

\section{Summary}
Prediction of the field penetration onset is an important subject to achieve stable and high-performance operation in tokamaks using EF correction and RMP-induced ELM control. In this Brief communication, we demonstrate the resonant layer response in the EV regime where not only the parallel but also curl elements of electron viscosity should be considered in the generalized Ohm's law. The field penetration scaling particularly implies that electron viscosity enhances the field penetration threshold in low rotation and high viscosity plasmas.

\begin{acknowledgments}
This work is supported by the Technology Development Projects for Leading Nuclear Fusion through the National Research Foundation of Korea (NRF) funded by the Ministry of Science and ICT (No. RS-2024-00281276).
\end{acknowledgments}

\appendix



\section*{Reference}
\bibliographystyle{unsrt}
\bibliography{references}

\end{document}